\documentclass[a4paper,fleqn,usenatbib]{mnras}

\usepackage{newtxtext,newtxmath}

\usepackage[T1]{fontenc}
\usepackage{ae,aecompl}

\usepackage{graphicx}	
\usepackage{amsmath}	
\usepackage{amssymb}	
\usepackage{aas_macros}
\usepackage{hyperref}

\title[A mock catalogue for the PAU Survey]{The PAU Survey: Spectral features and galaxy clustering using simulated narrow band photometry}

\author[L. Stothert et al.]{
L. Stothert$^{1,2}$,
P. Norberg$^{1,2}$,
C. M. Baugh$^{1}$,
A. Alarcon$^{3,4}$,
A. Amara$^{5}$, \newauthor
Jorge Carretero$^{6}$\thanks{Also at Port d'Informaci\'{o} Cient\'{i}fica (PIC), Campus UAB, C. Albareda s/n, 08193 Bellaterra (Cerdanyola del Vall\`{e}s), Spain},
F.J. Castander$^{3,4}$, 
M. Eriksen$^{6}$\footnotemark[1],
E. Fernandez$^{6}$,
P. Fosalba$^{3,4}$, \newauthor
J. Garcia-Bellido$^{7}$,
E. Gaztanaga$^{3,4}$, 
H. Hoekstra$^{8}$,
C. Padilla$^{6}$,
A. Refregier$^{5}$, \newauthor
E. Sanchez$^{9}$,
L. Tortorelli$^{5}$.
\\
$^{1}$Institute for Computational Cosmology, Department of Physics, University of Durham, South Road, Durham DH1 3LE, UK\\
$^{2}$Centre for Extragalactic Astronomy, Department of Physics, University of Durham, South Road, Durham DH1 3LE, UK\\
$^{3}$Institute of Space Sciences (ICE, CSIC), Campus UAB, Carrer de Can Magrans, s/n, 08193 Barcelona, Spain \\
$^{4}$Institut d'Estudis Espacials de Catalunya (IEEC), E-08034 Barcelona, Spain\\
$^{5}$Institute for Particle Physics and Astrophysics, ETH Z\"{u}rich, Wolfgang-Pauli-Str. 27, 8093 Z\"{u}rich, Switzerland\\
$^{6}$Institut de F\'{i}sica d'Altes Energies (IFAE), The Barcelona Institute of Science and Technology, Campus UAB, 08193 Bellaterra (Barcelona) \\
$^{7}$Instituto de Fisica Teorica UAM/CSIC, Universidad Autonoma de Madrid, Cantoblanco 28049 Madrid, Spain\\
$^{8}$Leiden Observatory, Leiden University, Niels Bohrweg 2, Leiden, The Netherlands\\
$^{9}$Centro de Investigaciones Energ\'{e}ticas, Medioambientales y Tecnol\'{o}gicas (CIEMAT), Avenida Complutense 40, E-28040, Madrid, Spain
}

\date{Accepted XXX. Received YYY; in original form ZZZ}

\pubyear{2018}

\begin{document}
\label{firstpage}
\pagerange{\pageref{firstpage}--\pageref{lastpage}}
\maketitle

\begin{abstract}
We present a mock catalogue for the Physics of the Accelerating Universe Survey (PAUS) and use it to quantify the competitiveness of the narrow band imaging for measuring spectral features and galaxy clustering. The mock agrees with observed number count and redshift distribution data. We demonstrate the importance of including emission lines in the narrow band fluxes. We show that PAUCam has sufficient resolution to measure the strength of the 4000\AA{} break to the nominal PAUS depth. We predict the evolution of a narrow band luminosity function and show how this can be affected by the OII emission line. We introduce new rest frame broad bands (UV and blue) that can be derived directly from the narrow band fluxes. We use these bands along with D4000 and redshift to define galaxy samples and provide predictions for galaxy clustering measurements. We show that systematic errors in the recovery of the projected clustering due to photometric redshift errors in PAUS are significantly smaller than the expected statistical errors. The galaxy clustering on two halo scales can be recovered quantatively without correction, and all qualitative trends seen in the one halo term are recovered. In this analysis mixing between samples reduces the expected contrast between the one halo clustering of red and blue galaxies and demonstrates the importance of a mock catalogue for interpreting galaxy clustering results. The mock catalogue is available on request at \url{https://cosmohub.pic.es/home}.
\end{abstract}




\section{Introduction}
Clustering measurements at low redshifts have been shown to display a dependence on galaxy properties such as stellar mass, luminosity, and colour, which suggests that these properties depend on the mass of the host dark matter halo (e.g. \citealt{norberg2dfclustering,zehavisdss}). Galaxy clustering measurements are therefore not only useful for constraining the cosmological model but also for developing our understanding of galaxy formation physics. 

The processes that shape how the efficiency of galaxy formation depend on halo mass may change with redshift, so it is important to extend measurements of galaxy clustering as a function of intrinsic galaxy  properties to higher redshift. One clear piece of evidence hinting at evolution in the galaxy formation process is the dramatic change in the amount of star formation activity since $z \sim 1 - 2$, with roughly ten times less star formation globally by the present day \citep{Lilly:1996,Madau:1996}. 

The measurement of clustering as a function of galaxy properties poses different challenges to those faced when using large-scale structure to constrain cosmological parameters. In the cosmological case, the aim is to maximize the volume probed whilst maintaining an appropriate number density of galaxies to achieve a moderate signal-to-noise ratio in the power spectrum measurement (e.g. \citealt{Feldman:1994}). The signal-to-noise ratio can be boosted by targeting galaxies with stronger clustering or a larger bias than the average population; beyond this, the selection of the galaxies is not that important in the cosmological case. On the other hand, when using clustering to probe galaxy formation, the desire is for a high number density of galaxies with a uniform selection covering a wide baseline in the intrinsic galaxy property of interest. 

Progress towards compiling large-scale structure samples for galaxy formation studies at intermediate redshifts has been made through the Galaxy And Mass Assembly Survey (GAMA; \citealt{gamadriver}), which targets galaxies in the $r$-band brighter than $r=19.8$, with a median redshift of $z \sim 0.2$ over 286 sq deg with high completeness, and the VIMOS Public Extragalactic Redshift Survey (VIPERS), which obtained redshifts for $86 \,7765$ galaxies with $i_{\rm AB} < 22.5$ over 24 deg$^2$ at $\sim$47\% completeness \citep{Scodeggio:2018}. The PRIsm MUlti-object Survey (PRIMUS; \citealt{primussurvey}) used slit masks to measure $\sim 2500$ redshifts in a single telescope pointing, recording $130 \,000$ redshifts over 9.1 deg$^{2}$ to $i_{\rm AB} = 23.5$, with a redshift distribution peaking at $z \sim 0.6$. These surveys have been used to carry out a large number of analyses to quantify the galaxy populations and to constrain the cosmological model. Below we highlight some results from these surveys which explicitly focus on using galaxy clustering measurements to probe the physics of galaxy formation. \cite{Farrow:2015} measured galaxy clustering as a function of luminosity and colour using GAMA. \cite{Loveday:2018} inferred the pairwise velocity distribution using the small scale galaxy clustering measured from GAMA. In both cases, these observational results were compared to theoretical models of the sort we will use here. \cite{Marulli:2013} used VIPERS to measure the dependence of galaxy clustering on stellar mass and luminosity for $0.5 < z < 1.1$. \cite{Coupon:2015} combined clustering measurements with a gravitational lensing analysis to constrain the galaxy halo connection.  \cite{Skibba:2014} measured the clustering of galaxies in PRIMUS as a function of colour and luminosity, \cite{Skibba:2015} studied the variation of the clustering amplitude with stellar mass and \cite{Bray:2015} examined how the luminosity dependence of clustering depends on pair separation. 

A limitation of spectroscopic surveys is the number of redshifts that can be measured in a single telescope pointing. This is set by the number of fibres or slits available to deploy to measure galaxy redshifts in the field of view. The use of some form of aperture to capture the light from a single galaxy also introduces a systematic effect on the clustering measured on small scales. The physical size of the slit or fibre means that in some cases only one member of a pair of galaxies within a particular angular separation can be targeted for a redshift measurement. This ``fibre collision'' effect can be mitigated by repeat observations of the same field or by applying a correction to the measured pair counts. 

An alternative to using spectroscopy to measure the radial distance to a galaxy is to use photometry taken in a number of bands. A photometric redshift can be assigned to a galaxy by, for example, comparing the observed flux in different bands to that derived from a template spectrum that is shifted in redshift \citep{Benitez:2000,Bolzonella:2000}. The photometric redshift approach has three advantages over spectroscopy: 1) the galaxy selection is homogeneous down to the flux limit, without any bias towards a higher success rate of redshift measurement for galaxies with emission lines (although the precision of photometric redshifts does depend on the colour of the galaxy; see e.g. \citealt{Marti:2014,Sanchez:2014}), 2) there are no `fibre collisions' that can impact galaxy clustering measurements and 3) there is no requirement to match the surface density galaxies to the number of slits or fibres within the field of view.

Broad band photometry, in which the typical filter width is $\sim 1000$ \AA, is limited to a redshift precision $\Delta z/(1+z)$ (hereafter $\sigma_z$) of $\sim$3-5\%. CFHTLS wide, a broad band survey observing in $u, g, r, i$ and $z$ which is 80\% complete to $i < 24.8$ reaches $\sigma_z \sim 3 \%$ for $i < 24$ with $\sim 4\%$ catastrophic errors (defined as $\sigma_z > 15 \%$) 
\citep{cfhtredshifts}. This level of precision is sufficient to divide galaxies into redshift shells in which the projected clustering can be measured. The error in the radial distance estimate in this case is $\sim 100 h^{-1} \,{\rm Mpc}$ at z = 0.7. 

The accuracy of photometric redshifts can be improved by using narrower filters \citep{Wolf:2004}. The Advanced Large, Homogeneous Area Medium Band Redshift Astronomical Survey (ALHAMBRA) \cite{alhambra}, offers a recent example of this by using 20 medium band filters, each $\sim$300\AA{} in width, to reach an accuracy of $\sigma_{\rm z}   = 1.4$\% for galaxies with $i < 24.5$ \citep{alhambraphotoz}. \cite{cosmosthirty} reached $\sigma_z = 1.2$\% for objects with $i < 24$ over the 2 deg$^{2}$ COSMOS field using a combination of broad, medium and narrow bands spanning the ultra-violet to the mid-infrared.

\begin{figure}
  \includegraphics[width=\linewidth]{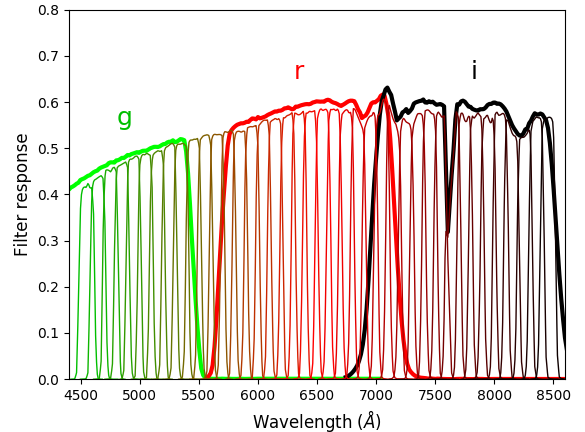}
  \caption{Filter response as a function of wavelength for the 40 PAUcam filters (thin lines) compared to CFHT MegaCam broad band filters
$g,r,i$ (thick lines). Filter response curves include atmospheric transmission, telescope optics and CCD quantum efficiency.}
  \label{fig:filters}
\end{figure}

The Physics of the Accelerating Universe Survey (PAUS) is a narrow band imaging survey using PAUCam, \cite{paucam}, which was commissioned in June 2015, on the $4.2\,$m William Herschel Telescope, and Padilla et al. (In prep). PAUS will measure narrow band fluxes by using forced photometry on objects previously detected in overlapping broad band photometric surveys CFHTLenS \citep{CFHTLenS} and KiDS \citep{kids}. PAUS aims to perform forced photometry measurements in 40 narrow bands over 100 deg$^2$ for objects $i < 23$, and reach signal-to-noise of 3 at narrow band magnitude 23. Each of the 40 narrow band filters have FWHM 130\AA{} and are spaced by 100\AA{}, over the wavelength range of $4500$\AA{} to $8500$\AA{} \citep{Marti:2014}. Fig.~\ref{fig:filters} shows the PAUCam narrow band filters compared to the $g, r$ and $i$ bands from CFHTLS. 40 narrow bands span the region covered by these three broad band filters. The increased spectral resolution of PAUS imaging will allow for photometric redshift measurements of $\sigma_{\rm z}$ = 0.35\% for objects $i < 23$ \citep{marti}. This represents an improvement of nearly an order of magnitude compared with typical broad band redshift measurement uncertainties, and in principle allows the radial distance information to be used in clustering estimates and to infer membership of galaxy groups.

The spectral features of a galaxy encode information about intrinsic properties such as its stellar mass, age and metallicity. Using these properties to define samples for clustering studies can then help us to understand the connection between galaxy properties and the mass of the host dark matter halo. These features include emission lines, absorption features, the 4000\AA{} break and the shape of the continuum.   Measuring the spectral features of individual galaxies has largely been in the domain of spectroscopic surveys. \cite{kauff} used a combination of the strength of the 4000\AA{} break and the H$\delta$ absorption feature to constrain the stellar age, and contribution to stellar mass from recent star formation events, for a large sample of galaxies drawn from the spectroscopic Sloan Digital Sky Survey. \cite{breakkband} used stacking to measure the average values of spectral features using the medium band photometry of 3500 galaxies from the NEWFIRM survey to constrain star formation histories 0.5 < z < 2.0. One of our goals here is to determine how competitively PAUS can be used to determine spectral features of galaxies, compared to the use of higher resolution spectra e.g. from zCOSMOS \citep{zcosmos}, allowing for any modifications to the definitions of the spectral features driven by the narrow band photometry and taking into account errors in the photometry and in the photometric redshift estimation.  

Here we use the galaxy formation model {\tt GALFORM} introduced by \cite{g13}, combined with a large-volume, high-resolution N-body simulation to build a mock catalogue for PAUS. \cite{Contreras:2013} demonstrated that semi-analytical models of galaxy formation give robust predictions for galaxy clustering and, where differences exist between the models, they can be traced back to choices made in the treatment of galaxy mergers and the spatial distribution of satellite galaxies (see also \citealt{Pujol:2017}). \cite{farrow} used the \cite{g13} model to interpret GAMA clustering measurements as a function of luminosity, stellar mass and redshift.

The layout of this paper is as follows. Section \ref{lightcone} introduces the galaxy formation model and the PAUS mock catalogue, Section~\ref{features} investigates the use of the PAUS narrow band filters to measure  galaxy spectral features, and Section~\ref{results} gives predictions for the narrow band luminosity functions, other characterisations of the galaxy population in PAUS and galaxy clustering. We conclude with Section~\ref{conclusion}.

\section{PAUS mock lightcone}
\label{lightcone}

Here we describe the N-body simulation and galaxy formation model used (\S~2.1), introduce  some basic properties of the mock catalogue constructed (\S~2.2), discuss the modelling of emission lines and their impact on narrow band fluxes (\S~2.3) and set out the treatment of errors in photometry and in photometric redshift errors.  

\subsection{N-body simulation \& galaxy formation model}

To model the galaxy population observed with PAUS we use the {\tt GALFORM} semi-analytic galaxy formation model presented in \cite{g13} (hereafter GP14). The {\tt GALFORM} model \citep{cole2000} aims to follow the formation and evolution of galaxies in dark matter halos by solving a set of differential equations that describe the transfer of mass and metals between reservoirs of hot gas, cold gas and stars (see the recent extensive description of the model by \citealt{Lacey:2016} and the reviews by \citealt{Baugh:2006} and \citealt{Benson:2010}). Due to the complexity and uncertainty of galaxy formation physics, many processes are modelled using equations which require parameter values to be specified. These are set by requiring the model to reproduce a selection of observations of the galaxy population, mostly at low redshift. The model calculates the star formation and merger history for each galaxy, including all of the resolved progenitors. With an assumption about the stellar initial mass function (IMF) and a choice of stellar populations synthesis (SPS) model, {\tt GALFORM} outputs the flux for each galaxy in the PAUS bands using the composite stellar population obtained from the star formation history \citep{GP:2013}. This includes a calculation of the attenuation in each band, based on the optical depth calculated from the metallicity of the gas and the size of the disk and bulge components of the galaxy \citep{GP:2013}   

To build a mock catalogue on an observer's past lightcone with spatial information about the model galaxies, it is necessary to implement the galaxy formation model in an N-body simulation. The dark matter halo merger trees used in the galaxy formation model are also extracted from the N-body simulation \citep{Jiang:2014}. The GP14 model is implemented in the Millennium WMAP7 N-body simulation (hereafter MR7, \citealt{guo2013}). The MR7 run has a halo mass resolution of $1.86 \times 10^{10} \, h^{-1} \, {\rm M_{\odot}}$ in a cube of side $ 500 h^{-1} {\rm Mpc} $. The use of the MR7 run means that the GP14 model is complete to $i < 23$ for $z > 0.2$. This is sufficient for our analysis. GP14 is an update of the model presented in \cite{lagos12} to make it compatible with the WMAP7 cosmology and includes the improved star formation treatment implemented by \cite{Lagos:2011}.

\subsection{Mock catalogue on the observer's past lightcone}

The depth of PAUS means that the properties and clustering of galaxies will evolve appreciably over the redshift range covered. Hence it is necessary to take this into account when constructing a mock catalogue for PAUS. The starting point is the galaxy population calculated using {\tt GALFORM} at each of the N-body simulation outputs. Following the lightcone interpolation described in \cite{mersonlc}, we construct a mock catalogue of one contiguous 60 sq deg patch. PAUS will target multiple fields but this will make little difference to one point statistics and small scale clustering results presented here. 

\begin{figure}
  \includegraphics[width=\linewidth]{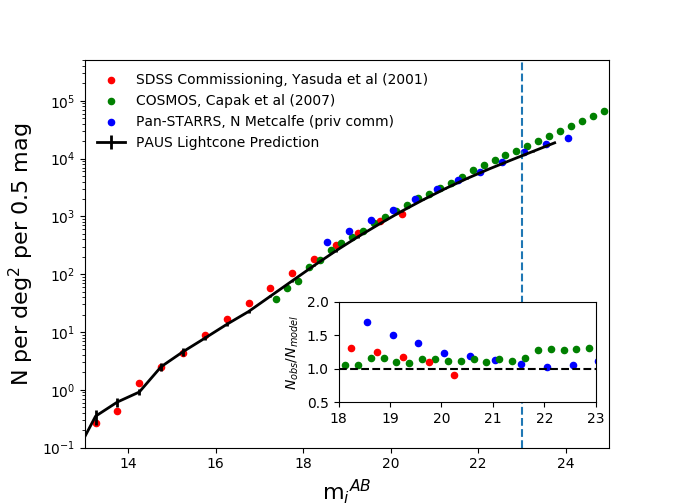}
  \caption{The predicted $i$-band galaxy number counts in the PAUS mock catalogue  (solid line) compared with various observations (coloured symbols; see legend). The vertical bars on the solid line show a jackknife estimate of the sample variance on the number counts. We have omitted the errors on the observational estimates of the counts as they come from very different solid angle surveys. The vertical blue dashed line indicates the PAUS magnitude limit $i = 23$. The inset shows, on a linear scale, the result of dividing the observed number counts by the lightcone predictions. }
  \label{fig:numbercounts}
\end{figure}

It is important to demonstrate that the mock catalogue is in broad agreement with the currently available  observational data. The number counts of the PAUS mock compare well with large area photometric surveys as shown by Fig.~\ref{fig:numbercounts}, which shows the agreement between the model and the observations from Pan-STARRS (N. Metcalfe, priv. comm) and the Sloan Digital Sky Survey \citep{sdss}. The systematic differences between the data points are partly due to the slightly different i band filters used in each survey. The offset between the mock catalogue and the data is reasonable when considering the systematic differences between the data. The low redshift incompleteness due to finite halo mass resolution of the WM7 simulation does not impact this comparison as the total number of faint objects is dominated by galaxies with $z > 0.2$, which are well resolved in the model.  

\begin{figure}
  \includegraphics[width=1.04\linewidth]{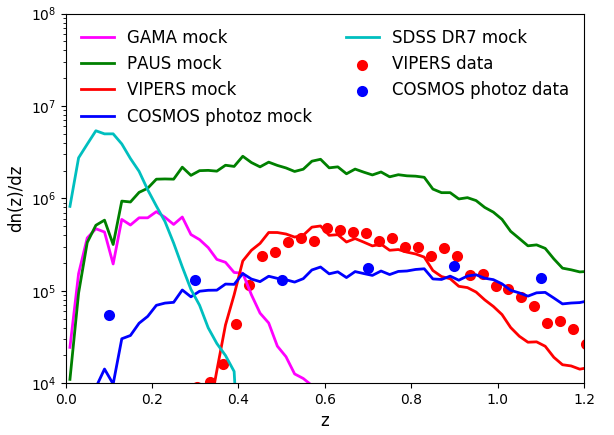}
  \caption{The redshift distributions in various mock catalogues (lines) compared to survey data (circles; see legend). The VIPERS data is taken from \protect\cite{vipersnz}, and the VIPERS mock catalogue is a 24 deg$^2$ lightcone to $i < 22.5$ with a 65\% sampling rate. The mock VIPERS n(z) is then statistically corrected for the colour cut using the empirical relation found in de la Torre et~al. The COSMOS photo-z data is taken from \protect\cite{cosmosthirty}, and the COSMOS photo-z mock is a 2 deg$^2$ lightcone retaining galaxies with $21.5 < i < 24.5$. The SDSS mock is a 10000 deg$^2$ lightcone with $r < 17.77$ and the GAMA lightcone covers 180 deg$^2$ to $r < 19.8$. These are plotted without an observational comparison to show the relative survey sizes and depths.}
  \label{fig:nz}
\end{figure}

Fig.~\ref{fig:nz} shows the redshift distributions for the mock lightcones associated with five different galaxy surveys, along with data from VIPERS \citep{vipersnz}, and COSMOS photo-z \citep{cosmosthirty}. The choice of the two comparison datasets was made to test the mocks against surveys with flux limits on either side of the nominal PAUS $i$-band magnitude limit, VIPERS $i < 22.5$ and COSMOS photo-z with $21.5 < i < 24.5$. The model predictions agree reasonably well with the observations. The disagreement with the lowest redshift COSMOS data point is due to incompleteness in the model; this will be less important for the PAUS mock which is shallower than the COSMOS one. There is some disagreement with the high redshift tail of the VIPERS n(z). This suggests that the model under predicts the bright end of the $i$-band luminosity function at higher redshifts. However, as our analysis is limited to $z < 0.9$, an investigation into the cause and significance of this discrepancy is left to a later date. For $z < 0.9$, the VIPERS mock catalogue  agrees well with the observations.

One current limitation of the mock catalogue is that it cannot be used for validation of photometric redshift codes. Tests run using the photo-z code embedded in the PAUS pipeline reveal discreteness in the returned redshifts which are aligned with MR7 snapshots. This issue arises due to the narrow width of the PAUS filters and the associated shift in redshift being smaller than the spacing of the N-body outputs in redshift. This is not an issue for broad band photometry or when using multiple adjacent filters for measurements as in this analysis. A catalogue constructed using the P-Millennium simulation (Baugh et al, in prep), will improve both the mass and time resolution of our lightcone mock catalogue.

\subsection{Impact of emission lines on narrow band fluxes}
\label{sec:impactemlines}

Emission lines are generally thought to make a negligible contribution to the flux measured in broad band filters, even for high redshift galaxies \citep{cowley}. However, the narrow width of the PAUcam filters means that it is necessary to revisit the contribution of emission lines for PAUS.

GALFORM makes a calculation of the emission line luminosity of each galaxy using the number of Lyman continuum photons, the metallicity of the star-forming gas and a model for HII regions from \cite{stasinska}. \cite{g17} give a recent illustration of this functionality presenting predictions for the abundance and clustering of OII emitters.  

\begin{figure}
  \includegraphics[trim={0.3cm 0 1cm 1cm},clip,width=\linewidth]{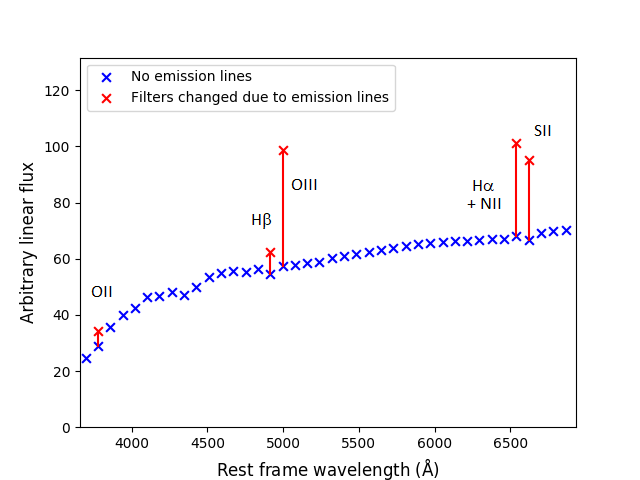}
  \caption{PAUCam filter fluxes for an illustrative star-forming galaxy taken from the PAUS mock. All 40 PAUCam filters are plotted. Blue (red) crosses show filter fluxes without (after including) emission lines.}
  \label{fig:emlines}
\end{figure}

Fig.~\ref{fig:emlines} shows the contribution emission lines can make to the PAUS narrow band fluxes for a single model galaxy. This illustrates that emission lines can be beneficial not only for the estimation of photometric redshifts, but suggests that PAUS could be used to identify and characterise populations of emission line galaxies. This is particularly relevant for the preparations for upcoming large spectroscopic surveys such as DESI \citep{desi} and Euclid \citep{euclid} which will build redshift catalogues from emission line galaxies.

\begin{figure}
  \includegraphics[width=\linewidth]{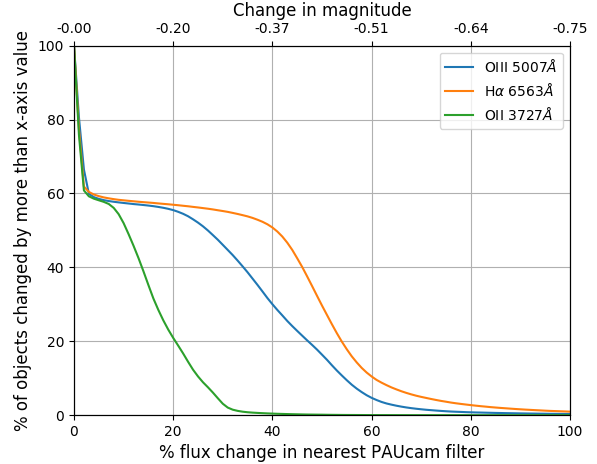}
  \caption{Fraction of model galaxies whose flux in nearest PAUCam filter is affected by the inclusion of a specific emission line (as indicated by the key). Only galaxies with redshift $0.21 < z < 0.3$ and magnitude $i < 23$ are shown to preserve a common sample where all lines can be sampled by a PAUCam filter. See section \ref{sec:impactemlines} for a discussion.}
  \label{fig:emlinesincatalogue}
\end{figure}

Fig.~\ref{fig:emlinesincatalogue} shows the fraction of galaxies whose relevant PAUS filter flux changes by a given percentage due to the contribution of one of the H$_{\alpha}$, OII or OIII emission lines. For this calculation we restrict ourselves to a redshift range over which all lines are visible in the PAUCam filter wavelength range  (see Table~\ref{tab:redshiftranges}). The curves show the change in the flux of the filter with peak transmission closest to the observed emission line. Note that as PAUS filters have a FWHM 130\AA{}, a full width of $\sim$ 135\AA{}, and are spaced by 100\AA{}, in a good fraction of cases a line will also contribute significantly to a second narrow band flux measurement.

It can be seen from Fig.~\ref{fig:emlinesincatalogue} that for 50\% of galaxies in this sample that at least one narrow band flux measurement changes by 40\% or more due to the inclusion of emission lines. That fraction falls to 38\% for OIII and to 5\% for OII, due to the average lower luminosity in these lines compared to that in the H$\alpha$ line.

\subsection{Photometry and redshift errors}
\label{errors}

Photometric redshift errors and photometry errors are added to the mock catalogue in post-processing. Two lightcones are produced, one with perfect photometry and correct redshifts and the other with PAUS-like errors applied.  These errors are defined as Gaussian redshift errors of $\sigma_z$ = 0.35$\%$ and Gaussian flux errors equivalent to a  signal-to-noise ratio of 3 at magnitude 23 in the narrow band filters. These redshift errors are a simple approximation to PAUS photo-z measurements which will be fully explored in Eriksen et al (in prep). No photometry errors are included in the broad band magnitudes as the sources of the broad band photometry will be at least one to two magnitudes deeper than the nominal depth of PAUS of $i = 23$.

\begin{figure*}
 \includegraphics[trim={0 2cm 0 1cm},clip,width=\linewidth]{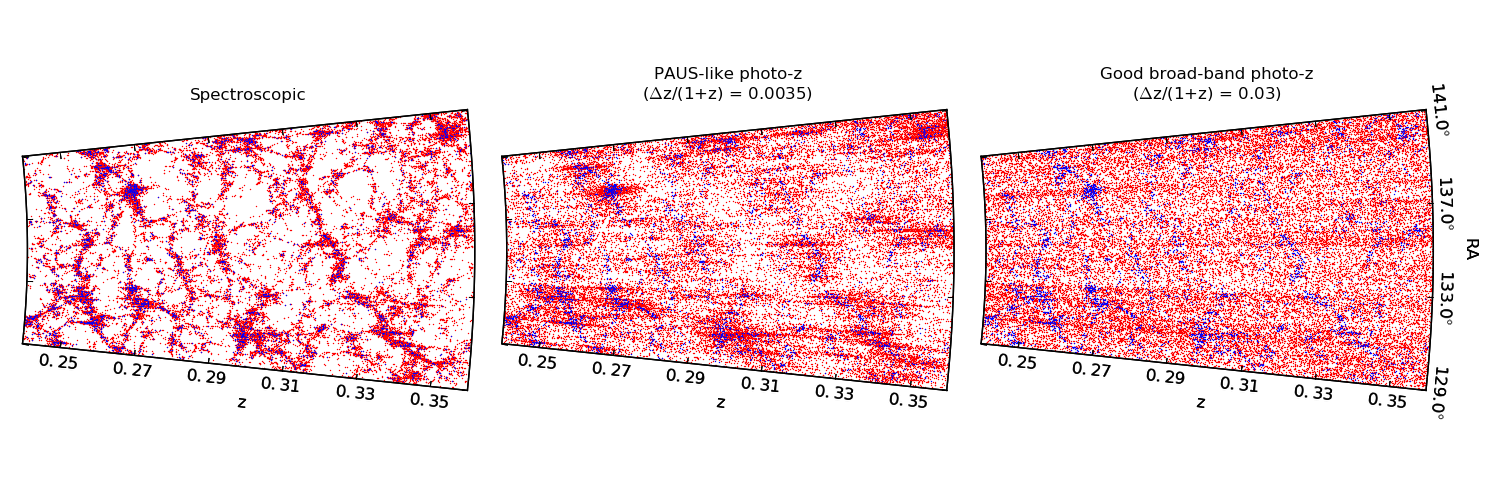}
  \caption{The spatial distribution of galaxies in a 1 degree thick slice from the PAUS mock catalogue. The three panels show the spatial distribution with spectroscopic redshift resolution (left), with PAUS-like redshift resolution (centre) and for typical broad band redshift resolution (right). Red points are galaxies brighter than the PAUS magnitude limit $i = 23$, while blue points correspond to GAMA galaxies ($r < 19.8$) with the spectroscopic redshift.}
  \label{fig:cone}
\end{figure*}

Fig.~\ref{fig:cone} shows the spatial distribution of galaxies in the PAUS mock catalogue and illustrates the impact of different redshift errors on the appearance of the large-scale structure of the universe traced by galaxies. Also shown in Fig.~\ref{fig:cone} are the model galaxies that satisfy the selection criterion for the GAMA survey, $r < 19.8$ (\citealt{gamadriver}; plotted at their spectroscopic redshift using blue points). The left panel of Fig.~\ref{fig:cone} highlights how much richer structures will be in a spectroscopic PAUS compared with GAMA, due to the deeper flux limit. The middle panel of Fig.~\ref{fig:cone} shows that a significant amount of radial information is retained once the redshifts of the mock galaxies are perturbed by the photometric redshift errors expected for PAUS. At $z \sim 0.3$, the expected photometric redshift errors for PAUS, $\sigma_z$ of 0.35\%, correspond to a comoving distance error of $\sim 13 h^{-1} \,{\rm Mpc}$. Hence, it will be feasible to extract information about group and cluster membership from PAUS (for an example of group finding in a catalogue with less accurate photometric redshifts than those expected in PAUS, see \citealt{Jian:2014}). The right panel shows how little radial position information is retained when applying the photometric errors expected for broad band photometry.    

\section{PAUS Galaxy properties}
\label{features}

The PAUS narrow band filters cover the wavelength range from 4500-8500\AA{} in which certain spectral features can be observed. Over the range in which PAUS will make the greatest contribution to clustering measurements, $0.25 < z < 1.0$, the rest frame wavelengths from 3000\AA{} to 4470\AA{} are always accessible with PAUS photometry. Table~\ref{tab:redshiftranges} lists the spectral features in the PAUS wavelength range that are investigated here. We assess the direct observation of these features given a galaxy with PAUS-like uncertainties in photometry and redshift. An alternative approach would be to extract the spectral information from the best fitting template spectral energy distribution to the PAUS fluxes, which is obtained as part of the photometric redshift estimation. Using the templates in this way could reduce the statistical error, as this approach uses information from all of the filters that are available for a given galaxy. However, this would introduce a systematic error through restricting the results to be derived from combinations of a limited number of templates. It will in fact be best to switch to using templates for measurements whose statistical errors exceed a certain threshold. The exact threshold is unknown as it depends on the unquantifiable systematic of template incompleteness, but this analysis can be used to define the point at which direct measurements become unfit for purpose, i.e when must we switch to using templates.

\subsection{Rest-frame defined broad bands}
\label{restframesection}

We define rest frame broad bands to best utilise the narrow band information from PAUS. These quantities are calculated by integrating the interpolated low resolution spectrum provided by the narrow bands. This type of direct rest frame measurement is possible because each of the PAUCam filters is flux calibrated, something which is often not the case with higher resolution spectra. 

\begin{figure}
\includegraphics[width=\linewidth]{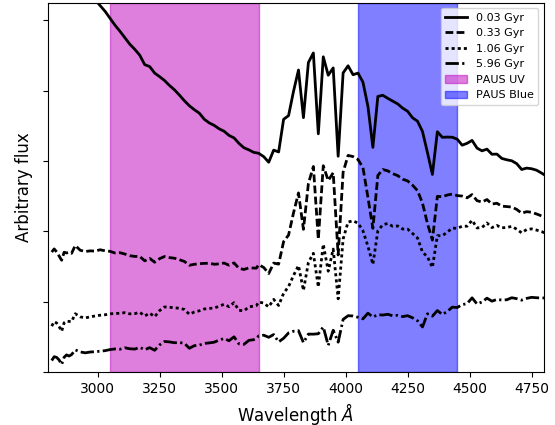}
\caption{The definition of new rest frame broad bands, PAUS UV (magenta) and PAUS Blue (blue). At $z = 0.6$, PAUS UV overlaps with 9.6 PAUCam filters and PAUS Blue overlaps with 6.4 PAUCam filters. The curves shown are some of the SEDs for single age stellar populations that are used in the construction of the mock catalogue. In all cases these are for one quarter solar metallicity, with ages given in the key.}
\label{fig:customfilters}
\end{figure}

As can be seen from Fig.~\ref{fig:customfilters} and Table~\ref{tab:redshiftranges}, the PAUS UV band has been chosen to be blue-wards of the 4000\AA{} break, and hence is sensitive to very young stars in the composite stellar population of a galaxy. Conversely, the PAUS Blue band is chosen to be red-wards of the break, and thereby probes somewhat older stellar content. PAUS UV is chosen to be wider than PAUS Blue to increase its signal-to-noise ratio. This is important particularly for the UV band due to the typical shape of an $i$-band selected galaxy SED meaning that, on average, the UV is fainter than the Blue. PAUS Blue can only be directly measured up to z = 0.9. 
  
  \begin{table}
  \begin{center}
 \begin{tabular}{ccc}
 \hline 
 Feature & Wavelength Range \AA{} & Redshift Range \\ 
 \hline
 OII  & 3727 & 0.21 - 1.28 \\
 OIII & 4959/5007 & 0.0 - 0.70 \\
 H$_\alpha$ & 6563 & 0.0 - 0.29 \\
 D4000$_\textrm{N}$ & 3850-3950 , 4000-4100 & 0.17 - 1.07 \\
 D4000$_\textrm{W}$ & 3750-3950 , 4050-4250 & 0.20 - 1.00 \\
 PAUS UV (M$_{\mathrm{UV}}^{\mathrm{h}}$) & 3050-3650 & 0.48 - 1.39 \\
 PAUS Blue (M$_{\mathrm{B}}^{\mathrm{h}}$) & 4050-4450 & 0.11 - 0.90 \\
 \hline
\end{tabular}
 \caption{Wavelength and redshift ranges over which PAUCam filters (4500-8500\AA{}) are sensitive to some common spectral features. The  table is limited to the main features observable over the redshift range $0.2 < z < 0.9$. See Fig.~\ref{fig:customfilters} for the definitions of the PAUS UV and PAUS Blue bands and see Fig.~\ref{fig:d4000exp} for the definitions of D4000. Note that M$^{\mathrm{h}}$ $\equiv$ M - 5log$_{10}$h.}
\label{tab:redshiftranges}
\end{center}
\end{table}

There are several benefits to using these new rest frame broad bands over and above single narrow bands or traditional broad bands:

\begin{itemize}
\item These bands cover multiple narrow band filters, increasing the signal-to-noise ratio of an individual measurement compared with using a single narrow band.
\item They are near direct measurements of galaxy rest frame SEDs and so do not require average $k$-corrections that broad band colour selections often require.
\item They can be chosen to sample desirable sections of a galaxy SED precisely.
\item Similar analyses can be performed on other photometrically calibrated spectra.
\item The filter wavelengths are fixed in the observer frame but sample a wavelength range in the rest frame that shrinks as $1/(1+z)$ with increasing redshift. This means that the rest frame magnitudes we have defined are measured using filters that become more closely spaced as the redshift of the source increases. Hence the rest frame magnitudes are better sampled with increasing redshift, which partly offsets the typical decrease in the signal-to-noise as sources get fainter.     
\end{itemize}




\begin{figure}
  \includegraphics[width=\linewidth]{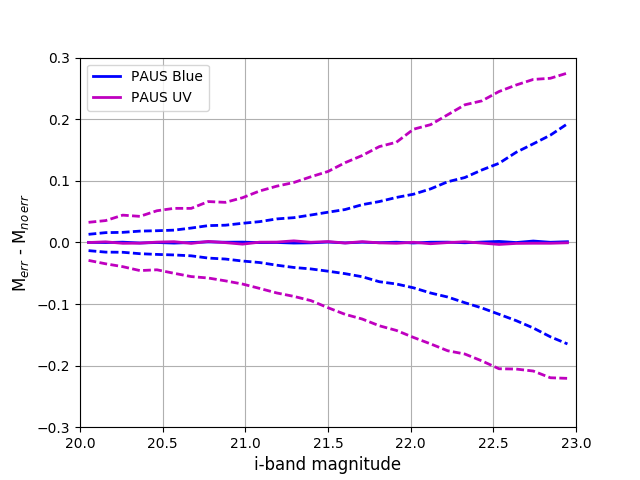}
  \caption{Statistical uncertainty in the PAUS UV and PAUS Blue magnitudes as a function of $i$-band magnitude for mock galaxies with $0.5 < z < 0.63$. The uncertainty includes redshift and photometry errors as described in Section~\ref{errors}. Solid lines show the median error and the dashed lines show the 10$-$90 percentile range.}
  \label{fig:magerrors}
\end{figure}

Fig.~\ref{fig:magerrors} shows how PAUS redshift and photometry errors propagate into errors in the PAU UV and PAU Blue magnitudes for a sample of mock galaxies with redshifts in the range $0.5 < z < 0.63$ and $i<23$. For 80 \% of model galaxies at $i = 23$ PAUS Blue can be measured to within $\pm$0.2 mags and PAUS UV to within $\pm$0.25 magnitudes. There is also no bias in the measurement at all values of i-band magnitude. Other redshift selections give similar errors and also show no bias.

\subsection{The 4000\AA{} break}
\label{subsection:d4000}

\begin{figure}
  \includegraphics[width=\linewidth]{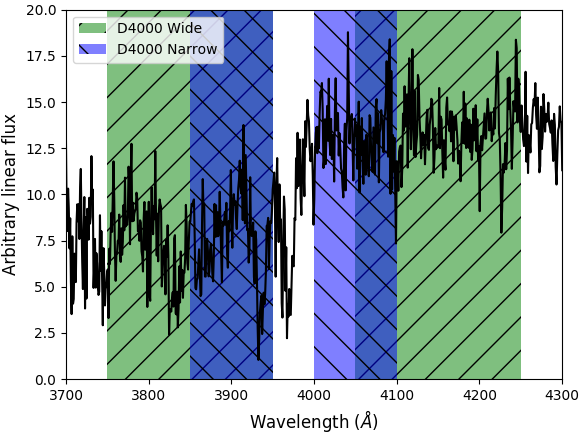}
  \caption{Definitions of D4000 wide and D4000 narrow overlaid on a randomly selected, de-redshifted, SDSS DR10 galaxy. The green shaded region represents the wide definition (3750-3950 and 4050-4250\AA{}) from \protect\cite{breakwide}, and the blue the narrow (3850-3950 and 4000-4100\AA{}) from \protect\cite{breaknarrow}.}
  \label{fig:d4000exp}
\end{figure}

The 4000{\AA} break is driven by a combination of CaII absorption lines and CN bands in the spectra of old stars. The quantity D4000 is the ratio of average flux in one spectral region at wavelengths just above 4000\AA{} and that in a region just below in wavelength. The literature defines this quantity in two ways, D4000 narrow defined in \cite{breaknarrow} and D4000 wide defined in \cite{breakwide}. The two flux bands used are different in each case and are visualised in Fig.~\ref{fig:d4000exp}. We first investigate if PAUCam has high enough resolution in a high signal to noise scenario to measure D4000 wide and narrow and then separately investigate D4000 measurements of PAUS mock galaxies.

\subsubsection{Measuring the 4000\AA{} break strength with PAUCam spectral resolution}

In order to test the measurement of the D4000 feature we look at a sample of 4500 SDSS DR12 galaxy spectra, selected around $z=0.1$ \citep{SDSSDR12, SDSSspectrograph}. We consider SDSS spectra for this test as the SPS used in {\tt GALFORM} are limited to 20\AA{} resolution. The SDSS galaxies were each randomly uniformally placed at a redshift in the range $0.2 < z < 0.9$ so that the different ways in which the PAUS filter can trace the feature are taken into account. The fluxes in the 40 PAUCam narrow bands were calculated for each galaxy. D4000 was then calculated using both the full resolution SDSS spectra, and then again by integrating a linear interpolation of the PAUS filter measurements. Both definitions of D4000 from the literature were calculated and results are presented with and without PAUS-like redshift errors, as defined in Section~\ref{errors}. We do not include photometry errors, as first we want to check if PAUCam has sufficient resolution to measure D4000 in a high signal to noise scenario.

\begin{figure}
  \includegraphics[width=\linewidth, trim={0 10 50 40},clip]{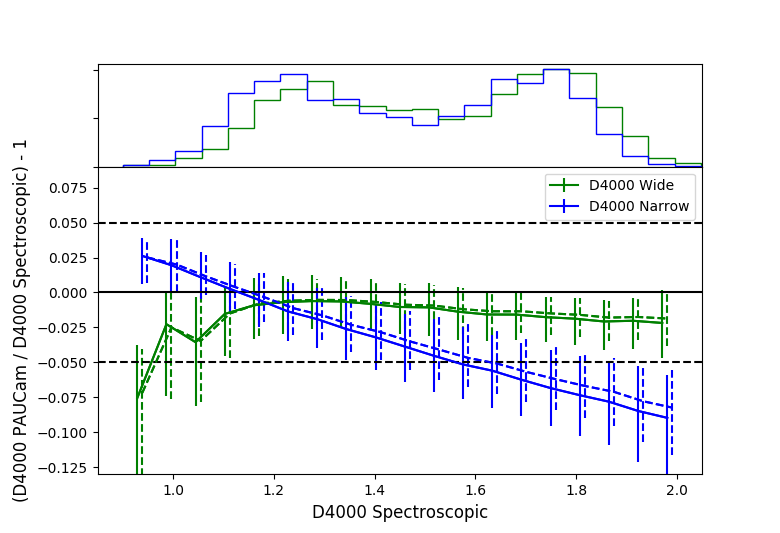}
  \caption{Relative accuracy with which D4000 can be recovered using PAUCam, as a function of the strength of D4000, measured using 4500 SDSS spectra observed at z $\sim$ 0.1 and redshifted over the interval $0.2 < z < 0.9$. D4000$_s$ is measured using the full spectra information while D4000$_P$ uses the PAUS filters. The green line shows the result for D4000 wide and the blue for D4000 narrow. Solid lines and error bars (which indicate the 10-90 percentile range) include a PAUS-like photo-z error while the dotted lines and error bars do not. Dotted lines are displaced in the x direction by 0.01 to make the error bars visible. The top panel shows the distributions of D4000 values for the sample.}
  \label{fig:d4000widevsnarrow}
\end{figure}

Fig.~\ref{fig:d4000widevsnarrow} shows how well interpolating between the PAUCam filters recovers the spectroscopic result for both the wide and narrow D4000 definitions from the literature. Both definitions of D4000 are biased due to the effective smoothing of a sharp spectral feature due to the finite width wavelength intervals used to calculate D4000. D4000$_n$ is affected by this bias more than D4000$_w$. The D4000$_n$ bias also scales as a function of the spectroscopic value for D4000 whereas the bias of D4000$_w$ is nearly constant with respect to this ideal. The D4000$_w$ measurement is biased by $\sim$2\%. This bias is not corrected for in later analysis as we will see in section \ref{d4000paus} that it is small compared to the random errors on PAUS mock galaxies. Once photometric redshift errors are included the error bars on both measurements increase only slightly. The error bars on D4000$_w$ are also smaller than those of D4000$_n$, $\sim\pm 2 \%$ and $\sim\pm 4 \%$ respectively. The superior recoverability of D4000$_w$ suggests this 4000\AA{} break definition should be used for PAUS measurements. The superior bias and noise performance of D4000$_w$ is to be expected as it overlaps with more PAUCam filters than D4000$_n$ does at a given redshift.

The redshift dependence of the D4000$_w$ measurement bias was investigated, as at each redshift the filters will trace the break in a different manner. The extreme scenarios are that the D4000 break lies mid-way across a filter or exactly in between two filters. It was found that the bias of D4000$_w$ varies by less than 1\% as a function of redshift. It is therefore not necessary to model this redshift dependence.


\subsubsection{4000\AA{} break strength in PAUS}
\label{d4000paus}

\begin{figure}
  \includegraphics[width=\linewidth]{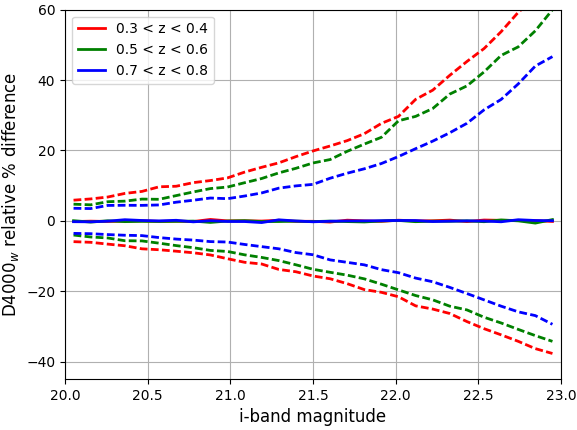}
  \caption{Relative percentage difference in D4000$_w$ as a function of $i$-band magnitude for different redshift slices. The relative percentage difference is defined as $100 \times (D4000_{\rm err} - D4000_{\rm true})/D4000_{\rm true}$, where the subscript err(true) refers to measurements made in the catalogue with(without) PAUS simulated redshift and photometric errors.}
  \label{fig:d4000catalogueerrors}
\end{figure}

To investigate the ability of using the PAUS photometry to measure D4000$_w$, this quantity is measured in both the mock catalogue with no errors and in the one with redshift and photometric errors introduced in section \ref{errors}. Fig.~\ref{fig:d4000catalogueerrors} shows the relative error in D4000$_w$ for redshift slices as a function of $i$-band magnitude. 80 percent of galaxies at $i = 23$ lie within 50\% of the true value of D4000$_w$. Photometric uncertainty is therefore the dominant source of error for PAUS galaxies. Looking at the population histogram in Fig.~\ref{fig:d4000widevsnarrow} it can be seen that the majority of galaxies have values of D4000$_w$  between $1.0$ and $2.0$, with a bimodal distribution peaking at 1.2 and 1.75. An error of 50\% is therefore very large compared to the range of D4000$_w$ . Galaxies with $i = 21.5$ and $z = 0.55$, however, are expected to have just a 15\% error in D4000$_w$, showing that direct D4000$_w$ measurements for a bright subset of PAUS objects are feasible. D4000$_w$ errors are smaller for higher redshift galaxies at a fixed $i$-band magnitude as the rest frame defined D4000$_w$ bands overlap with more PAUCam filters in this case than at lower redshifts. Individual studies will need to define a tolerable error for this quantity. Bimodal population cuts for example will be able to use a large subset of data and retain completeness and purity, whereas studies on the ages of individual galaxies may need to use a significantly restricted subset of the catalogue. One could also stack populations of galaxies and make a measurement on a mean spectra to reduce statistical error.


\section{Results}
\label{results}

In this section we review various properties of the galaxy population that we expect PAUS will be able to measure based on the predictions made using our mock catalogues. 

\subsection{Narrow band luminosity functions}
\label{narrowbandlfs}

The parameters in the {\tt GALFORM} model are calibrated to match low redshift observations, which are mainly one-point statistics such as the luminosity function. One of the applications of PAUS is to provide improved constraints on the model parameters by providing measurements of the narrow band luminosity function over a significant baseline in redshift.  

\begin{figure}
  \includegraphics[width=\linewidth, trim={10 0 50 40},clip]{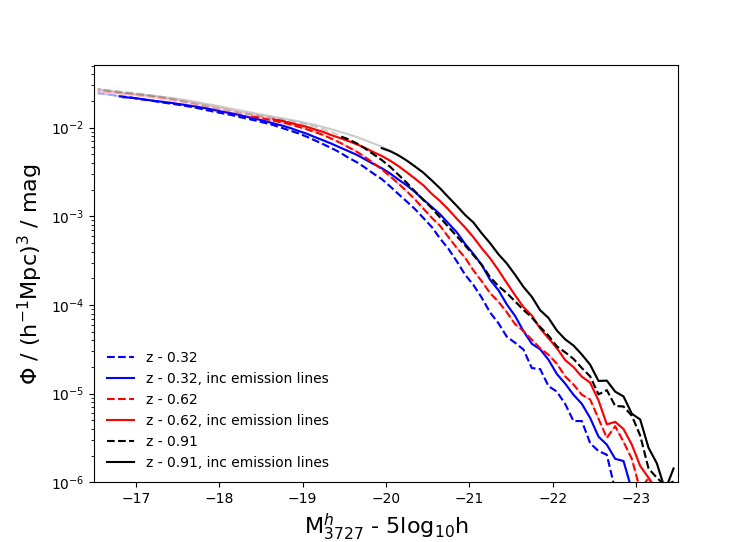}
  \caption{Luminosity functions at several snapshot redshifts (as labelled) of a PAUS filter at rest frame wavelength of  3727\AA{} $\pm$ 62.5\AA{}. A different PAUS filter is used at each redshift, chosen to overlap with the OII emission line. Solid lines show the prediction including the emission line flux and dashed lines do not. The plotted curves become fainter when they fall below 95\%  completeness at $i < 23$.}
  \label{fig:narrowbandlfs}
\end{figure}

We have seen that individual PAUCam narrow band magnitudes can be significantly affected by the emission line flux from a galaxy, so here we investigate the sensitivity of the narrow band luminosity functions to the inclusion of emission lines in the GP14 model (see \citealt{g17} for a further discussion of model predictions for OII emitters). Fig.~\ref{fig:narrowbandlfs} shows how a narrow band luminosity function of PAUCam like filter chosen to overlap in the rest frame with the OII emission line changes when the flux from the line is included. Measurements are made in the simulation snapshots. Inference of this quantity from observer frame measurements would require accurate $k$-corrections and is beyond the scope of this paper. It can be seen that neither redshift evolution nor inclusion of emission line flux change the faint end slope of the luminosity function in the GP14 model. The value of $M^{*}$, however, increases with both redshift, as a result of the increasing star formation, and also with the inclusion of OII line flux. The contribution of the stellar continuum to the flux in this band can be estimated by averaging the flux in bands placed at either side of the band that contains the OII emission, providing a constraint on the amount of emission line flux and its evolution with redshift.

\subsection{Characterisation of the galaxy population}
\label{colourdef}

\begin{figure*}
 \includegraphics[width=\linewidth]{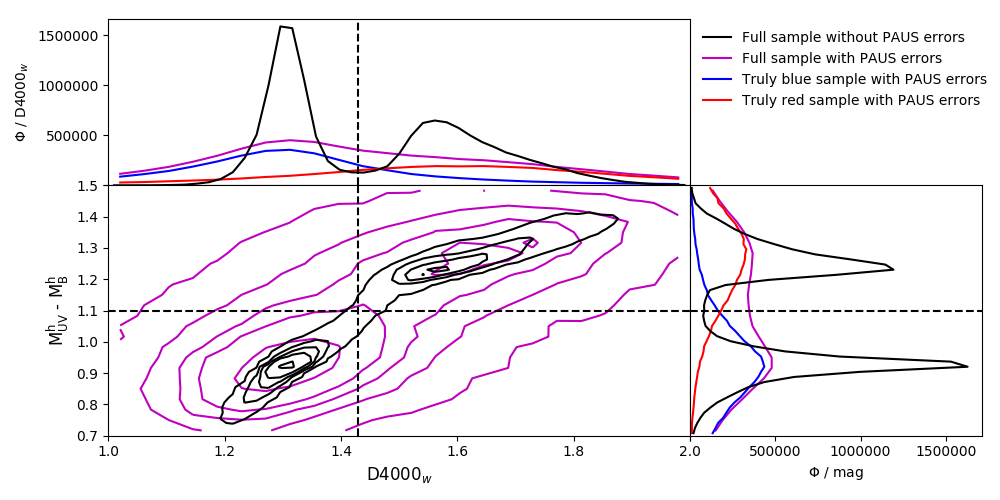}
  \caption{Distribution of galaxies with $i < 23$ and $0.5 < z < 0.63$ in the D4000$_w$ and M$_{\mathrm{UV}}^{\mathrm{h}}$ - M$_{\mathrm{B}}^{\mathrm{h}}$ colour plane, with and without simulated PAUS errors. The contours contain 10, 30, 50, 70 and 90\% of the sample. The solid black lines show the distributions for the full sample without errors and the magenta ones show the full sample with errors. The red (blue) curves show the distribution of galaxies that are intrinsically red (blue) in each measure when errors are included.}
  \label{fig:colourvsd4000}
\end{figure*}

One desirable objective for studying the evolution of the galaxy population is the ability to separate galaxies by colour in a consistent way across the redshift range sampled by PAUS. This objective can be achieved by using a cut in D4000 at $z < 0.5$ and a cut in M$_{\mathrm{UV}}^{\mathrm{h}}$ - M$_{\mathrm{B}}^{\mathrm{h}}$ above redshift 0.5. We could define a band further into the red to make a colour cut at lower redshifts, as M$_{\mathrm{UV}}^{\mathrm{h}}$ cannot be defined below z $\sim$ 0.5, see Table \ref{tab:redshiftranges}, but a cut in a different section of a galaxy SED may non-trivially select galaxies differently than the M$_{\mathrm{UV}}^{\mathrm{h}}$ - M$_{\mathrm{B}}^{\mathrm{h}}$ cut. In particular, the use of a redder colour selection might mix galaxies with different recent star formation histories, making clustering comparisons across redshift ranges less informative. The use of D4000$_w$ means that we are making a colour cut centred on the same portion of the SED as a cut in the colour M$_{\mathrm{UV}}^{\mathrm{h}}$ - M$_{\mathrm{B}}^{\mathrm{h}}$.

Fig.~\ref{fig:colourvsd4000} shows the distributions of D4000$_w$ and M$_{\mathrm{UV}}^{\mathrm{h}}$ - M$_{\mathrm{B}}^{\mathrm{h}}$ for a redshift range in which both can be measured. Both quantities show a bimodal distribution, which we can loosely refer to as `red' and `blue' populations. A cut is made at D4000$_w$ = 1.42 and M$_{\mathrm{UV}}^{\mathrm{h}}$ - M$_{\mathrm{B}}^{\mathrm{h}}$ = 1.1. Before photometric errors are added, disagreements in red-blue classification when using the two measures are at the sub-percent level. The cut in M$_{\mathrm{UV}}^{\mathrm{h}}$ - M$_{\mathrm{B}}^{\mathrm{h}}$ is appropriate to split the bimodal population at higher redshifts, as is the cut in D4000$_w$ for lower redshifts. Comparisons carried out using the model rest frame bands show that these colour cuts are similar to a traditional broad band rest frame cut in $u - g$. When including photometric errors, mixing between the red and blue populations is more severe when using D4000 than with the rest frame magnitudes due to the larger fractional error in D4000$_w$ at a fixed $i$ band magnitude (see Sections~\ref{restframesection} and \ref{d4000paus}). Errors on the M$_{\mathrm{UV}}^{\mathrm{h}}$ - M$_{\mathrm{B}}^{\mathrm{h}}$ colour are driven largely by errors in the UV magnitude.

\subsection{Galaxy clustering}

We select volume limited galaxy samples for clustering measurements based on redshift, PAUS blue luminosity and rest frame colour (as defined in Section~\ref{colourdef}). We choose not to split samples based on inferred quantities such as star formation rate or stellar mass as the inference of these properties from narrow band photometry is left to future work. Inferring these properties has also been shown to introduce biases based on the assumptions made in these inferences \citep{mitchell:2013}. In the mock including simulated PAUS errors the cuts are made after all sources of error are included. See Appendix \ref{app:code} for clustering definitions, details of the calculations and open source code links, and Appendix \ref{app:samples} for more information on sample selection. All errors in this section are calculated by using a jackknife over 12 regions in the simulated survey, see e.g \cite{Norberg:2009}. 

We estimate the galaxy bias from the ratio of the projected galaxy clustering to the projected clustering of the MR7 dark matter at the median redshift of the sample in question. The values of the correlation function for the MR7 snapshots were taken from \cite{nualapaper}. This quantity allows us to separate the evolution of the dark matter over time from the evolution of the galaxy population. On large scales this quantity is equal to the linear bias. More specifically we define projected galaxy bias as

\begin{equation}
\label{eq:bias}
b(r_p, z)= \sqrt{\frac{w_p(r_p, z)}{w_p(r_p, z)_{DM}}},
\end{equation}

where $w_p(r_p, z)$ is the projected correlation function defined in Eqn.~\ref{eq:wprp}.

\subsubsection{Impact of photometric uncertainty}

\begin{figure}
  \includegraphics[width=\linewidth, trim={20 0 50 40},clip]{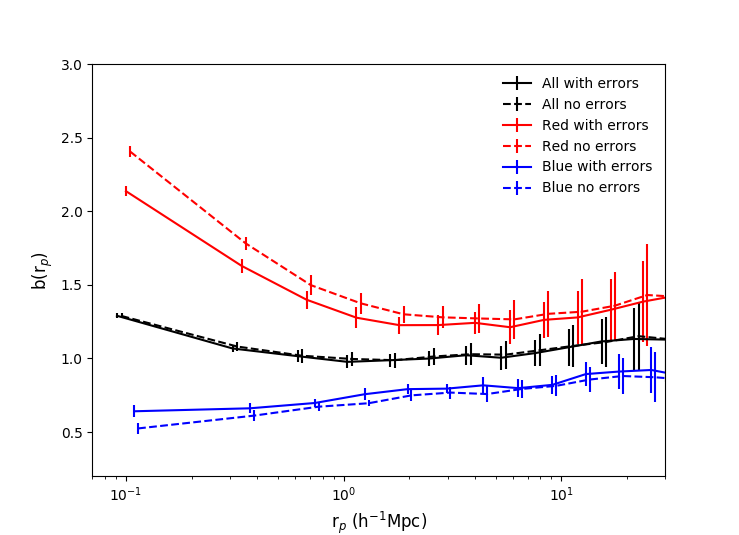}
  \caption{Projected galaxy bias (Eqn.~\ref{eq:bias}) for a typical PAUS sample ($0.5 < z < 0.63$, $-19.5 < \mathrm{M}_{\rm B}^{\mathrm{h}} < -19.0$). The full galaxy sample is shown in black and the results on splitting the sample into red and blue populations are shown in these colours. Solid lines show the results using the lightcone with redshift and photometry errors taken into account and the dashed lines the results without including these uncertainties. Errors calculated using jackknife resampling.}
  \label{fig:wprprecovery}
\end{figure}

Fig.~\ref{fig:wprprecovery} shows the bias measured for one mock PAUS sample (-$19.5 < \mathrm{M}_{\mathrm{B}}^{\mathrm{h}} < -19.0$) in the redshift range $0.5 < z <  0.63$, both with and without PAUS magnitude and photometric redshift errors. The value of $\pi_{\rm max}$ used was $ 100 h^{-1} {\rm Mpc}$. Fig. \ref{fig:wprpzerr} in the Appendix shows the recovery of the projected correlation as a function of different photometric redshift errors. A value of $\pi_{\rm max}$ of $ 50 h^{-1} {\rm Mpc}$ would have been sufficient for the photometric redshift errors assumed in this work, and would have slightly reduced the statistical noise, but the real survey will have a distribution of photometric redshift errors so the conservative value of $ 100 h^{-1} {\rm Mpc}$ was chosen.

For the sample selected only on redshift and M$_{\mathrm{B}}^{\mathrm{h}}$, the black lines in Fig. \ref{fig:wprprecovery}, the projected clustering signal is recovered without systematic error when including PAUS-like errors. The jackknife statistical errors only slightly increase when compared with the ideal case. This demonstrates that the PAUS photo-z measurements are sufficient to calculate the projected galaxy clustering without systematic error. Table \ref{tab:samplesnocolour} in the appendix shows that the sample with PAUS-like errors is over 90\% pure and complete. For the same sample with photo-z errors only and no photometry error these numbers both rise above 96\%, showing that mixing between samples due to photometric redshift errors is minimal.

Once a colour cut is applied to the full magnitude limited galaxy sample, a significant difference can be seen in the projected bias measurements for the red and blue populations. Errors in the photometry introduce mixing between the red and blue populations which leads to a small reduction in the difference between the one-halo scale projected bias of red and blue galaxies. Nevertheless the difference between the clustering measurements for these populations remains significant. Systematics on two-halo scales are within the statistical uncertainties. This confirms that the most significant source of systematic error in this analysis will be one one-halo scales and come from the misclassification of galaxies into red or blue sub-samples using these direct rest frame measurements. This systematic error shows up here as there is a large contrast between the one-halo clustering of red and blue samples, and PAUS will have small statistical errors on those scales. Again, statistical colour errors could be reduced by using the best fit photo-z SED inferred colours for fainter samples, but this is not tested here. This highlights the importance of understanding sample selection and the role of mock catalogues in interpreting clustering results.

\subsubsection{The redshift evolution of clustering}

\begin{figure}
  \includegraphics[width=\linewidth, trim={15 0 50 40},clip]{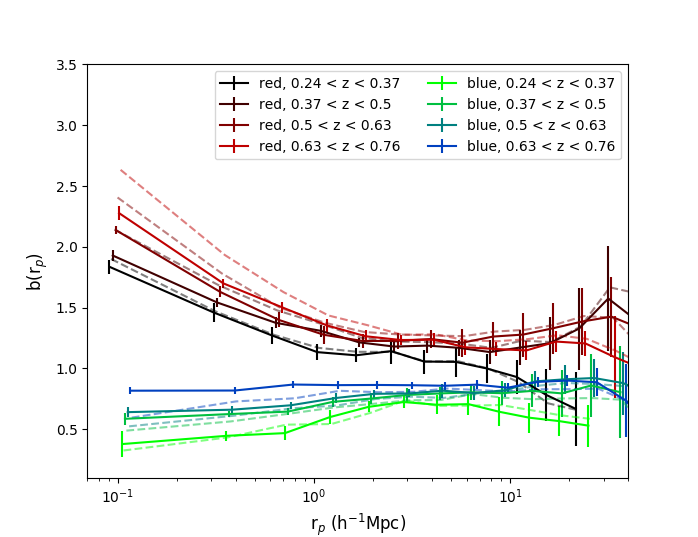}
  \caption{Projected galaxy bias (Eqn.~\ref{eq:bias}) inferred from the projected correlation function measured for samples with $-19.5 < \mathrm{M}_{\rm B}^{\mathrm{h}} < -19.0$, split by colour and redshift. Solid lines show the results using the lightcone including redshift and photometry errors and the dashed lines show the results
without these uncertainties. Errors, from jackknife resampling, only shown for PAUS-like sample.}
  \label{fig:bias}
\end{figure}

Fig.~\ref{fig:bias} shows the predicted redshift evolution of projected galaxy bias measured for samples of red and blue galaxies with $-19.5 < \mathrm{M}_{\rm B}^{\mathrm{h}} < -19.0$. Our estimate of the bias naturally takes into account the evolution of the clustering of the dark matter over this redshift interval. For all redshift bins red galaxies show stronger clustering than blue galaxies. This difference becomes larger for pair separations below $\sim 1 h^{-1} {\rm Mpc}$ corresponding to pairs within common dark matter halos. The bias also increases with redshift for both red and blue samples. This trend is also seen in all the other luminosity bins we have explored. This result, the decline in the bias as the universe ages, is due to faster growth of the dark matter correlation function compared with that of the galaxy correlation function over the same period, see e.g. \cite{Baugh:1999}. Again the systematic errors on two-halo scales are within statistical uncertainties. Qualitative trends seen on one-halo scales are preserved once errors are included, but the contrast between red and blue one-halo clustering is reduced due to colour mixing.

\subsubsection{The luminosity dependence of galaxy clustering}

\begin{figure}
  \includegraphics[width=\linewidth, trim={20 0 50 40},clip]{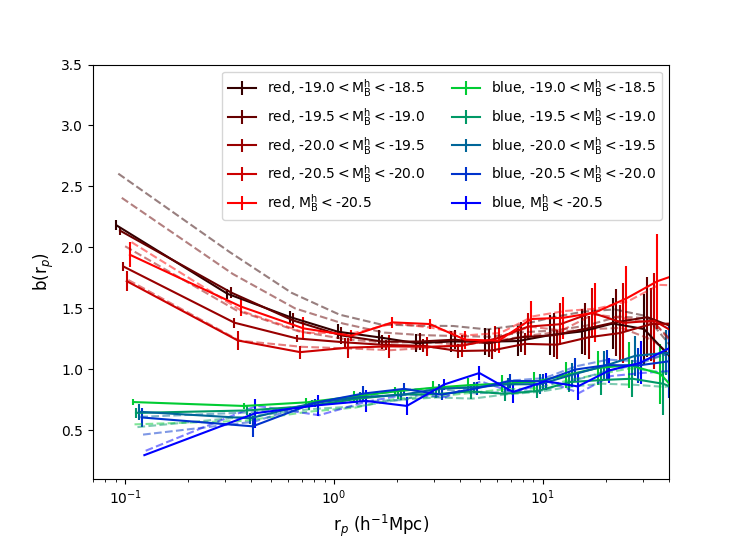}
  \caption{Projected galaxy bias (Eqn.~\ref{eq:bias}) inferred from the projected clustering measured for samples $0.5 < z < 0.63$, split by colour and M$_{\rm B}^{\mathrm{h}}$. Line types as in Figure \ref{fig:bias}.}
  \label{fig:biasoverlum}
\end{figure}

Fig.~\ref{fig:biasoverlum} shows the model prediction for the luminosity dependence of galaxy clustering. The split between the red and blue galaxies is once again very evident. As commented above, the red samples have stronger clustering than their blue counterparts. There is little luminosity dependence of the clustering measure for the blue samples (see also \citealt{Kim:2009} for a discussion of the luminosity dependence of clustering in an earlier version of the {\tt GALFORM} model used here). On the other hand, the clustering of the red samples shows a moderate dependence on luminosity which weakens on large scales and does not preserve the same ordering with luminosity that is displayed on small scales. Once again two-halo scale results are recovered within statistical errors.

\begin{figure}
  \includegraphics[width=\linewidth]{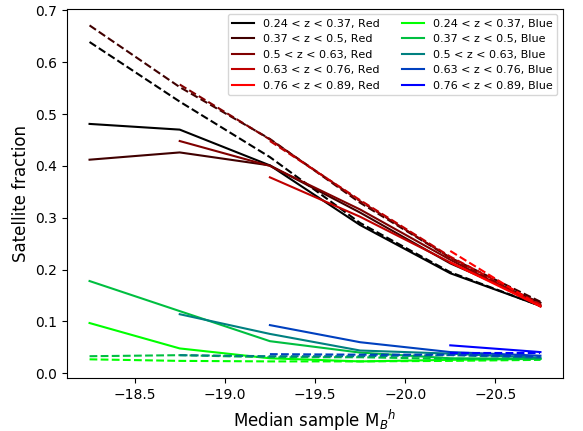}
  \caption{Satellite fraction as function of M$_{\rm B}^{\mathrm{h}}$ for galaxy samples split by colour and redshift. Line types as in Figure \ref{fig:bias}.}
  \label{fig:satfrac}
\end{figure}

One reason for the inverted trend of clustering decreasing with luminosity seen on small scales is due to the dominance of satellite galaxies in the lower luminosity red samples. This can be seen in Fig.~\ref{fig:satfrac}, which shows the satellite fractions of the clustering samples (Number of galaxies with satellite label in a sample divided by the total number of galaxies in the sample). Note that measuring this with the data would require significant modeling work. This figure also illustrates the impact of colour mixing on the satellite fraction of the samples. The lower luminosity bins at the lowest redshifts are significantly affected by mixing between central and satellites. These lower luminosity and redshift samples have the largest difference in satellite fraction between the red and blue populations and are the most likely to be misclassified in colour. This mixing error will either need to be modeled using mocks or we will have to rely instead on inferred colours extracted from an SED template, allowing for template incompleteness as a systematic error.

\section{Conclusions}
\label{conclusion}

We have introduced a mock catalogue built from a semi-analytical model of galaxy formation implemented in an N-body simulation for use in conjunction with the Physics of the Accelerating Universe Survey (PAUS). PAUS is a novel narrow band imaging survey which is underway on the William Herschel Telescope. The width of the PAUS filters means that photometric redshifts of unprecedented accuracy will become available for a homogeneously selected sample of galaxies down to $i =23$. The PAUS mock is built using the GP14 {\tt GALFORM} model \citep{g13}, which is run on the MR7 N-body simulation \citep{guo2013}. The galaxy snapshots produced at the output times of the MR7 run are then used to construct a mock catalogue on an observer's past lightcone, which predicts the evolution of the clustering of galaxies and their properties \citep{mersonlc}. The mock catalogue is available on request at CosmoHub\footnote{ \url{https://cosmohub.pic.es/home} } \citep{cosmohub}.

The resulting mock catalogue agrees with observed galaxy number counts to within the scatter between different surveys. Over the redshift range in which PAUS is expected to make the largest impact, $0.2 < z < 0.9$, the mock is in good agreement with the redshift distributions from COSMOS photo-z and VIPERS. There is some tension at $z > 1$ where the mock under predicts the VIPERS n(z), but this redshift range is less relevant for PAUS, and the observational errors are large at these redshifts \citep{vipersnz}. 

We include galaxy emission lines in the predicted PAUS measurements and show that this has a significant effect on PAUS narrow band fluxes. We show how the rest-frame narrow band luminosity function changes when emission lines are included by choosing a rest frame narrow band that overlaps with the OII emission line. The GP14 {\tt GALFORM} model predicts no change in the faint end slope of the narrow band luminosity function with or without emission line flux included and as a function of redshift. It does, however, predict an increase in M* with both redshift and on the inclusion of emission lines.

We define rest frame broad bands calculated directly from narrow band fluxes and predict that a PAUS Blue (PAUS UV) flux can be directly measured with an error of $\pm$ 0.15 ($\pm$0.25 mags) down to $i = 22.5$. These provide rest-frame measurements without needing to make any of the assumptions that come with average k-corrections used with broad band measurements. These rest-frame measurements are only possible because the PAUS narrow band measurements are flux calibrated. We show that the PAUCam filter set has sufficient resolution to measure the strength of the 4000\AA{} break, D4000. We predict that D4000$_w$ can be directly measured in PAUS to better than $\pm\sim$ 10\% precision for galaxies with $i < 21.5$. Providing errors on these quantities as a function of $i$-band magnitude will allow the PAUS data analysis pipeline to decide when to switch from directly measuring a quantity using the observed PAUCam filters to integrating over the best fitting SED assigned by a photometric redshift code. The latter incorporates statistical information from all filters but restricts results to a linear combination of SED templates, and is not explored here.

We explore galaxy clustering measurements over a redshift range of 0.2 to 0.9 for multiple luminosities and colours using the rest frame colours, D4000$_w$ and redshift. PAUS will provide a unique sample spanning this redshift range over a larger area than previously possible, with nearly 100\% completeness. No close galaxy pairs are missed as is often the case in spectroscopic surveys.

We show that systematic errors in projected clustering recovery due to PAUS photometric redshift errors are significantly smaller than statistical errors. All two-halo scale projected clustering results are recovered within statistical errors once PAUS redshift and photometry errors are included. One-halo scale clustering shows the same qualitative trends as measurements made in the ideal case but there is a loss of contrast between the one-halo scale clustering of red and blue galaxies caused by colour misclassification. This demonstrates the importance of a mock catalogue to interpret galaxy clustering results, particularly in the case of PAUS results on small scales, where statistical errors are small and any systematics are likely to be the dominant source of error.

We provide testable predictions for the mock catalogue that the measured galaxy clustering will evolve more slowly with redshift than the redshift evolution in the dark matter, especially for the one-halo term. The mock also predicts that red galaxies will cluster more strongly than blue galaxies. We also predict that fainter galaxies will cluster more strongly than brighter galaxies on small scales due to their larger satellite fraction, and that this trend will be particularly strong for red galaxies.

This work provides a tantalising illustration of the science that will be possible with PAUS, particularly with a view to constraining the galaxy - dark matter halo connection.

\section*{Acknowledgements} 

We would like to thank Nuala McCullagh for providing the MR7 dark matter correlation functions, Nigel Metcalfe for providing Pan-STARRS galaxy number counts, Violeta Gonzalez-Perez for providing the Galform model used in this work and John Helly for his help with the lightcone.

This work was supported by the Science and Technology Facilities Council 
[ST/J501013/1, ST/L00075X/1]. 
PN acknowledges the support of the Royal Society through the award
of a University Research Fellowship and the European Research
Council, through receipt of a Starting Grant (DEGAS-259586).
We acknowledge support from the Royal Society international exchange programme.
This work used the DiRAC Data Centric system at Durham University,
operated by the Institute for Computational Cosmology on behalf of the
STFC DiRAC HPC Facility \url{www.dirac.ac.uk}. This equipment was funded by
BIS National E-infrastructure cap- ital grant ST/K00042X/1, STFC
capital grant ST/H008519/1, and STFC DiRAC Operations grant
ST/K003267/1 and Durham University. DiRAC is part of the National E-Infrastructure.

Funding for PAUS has been provided by Durham University (via the ERC StG DEGAS-259586), ETH Zurich, Leiden University (via ERC StG ADULT-279396 and Netherlands Organisation for Scientific Research (NWO) Vici grant 639.043.512) and University College London. The PAUS participants from Spanish institutions are partially supported by MINECO under grants CSD2007-00060, AYA2015-71825, ESP2015-88861, FPA2015-68048, SEV-2016-0588, SEV-2016-0597, and MDM-2015-0509, some of which include ERDF funds from the European Union. IEEC and IFAE are partially funded by the CERCA program of the Generalitat de Catalunya. The PAU data center is hosted by the Port d'Informaci\'{o} Cient\'{i}fica (PIC), maintained through a collaboration of CIEMAT and IFAE, with additional support from Universitat Aut\`{o}noma de Barcelona and ERDF.

This work has made use of CosmoHub. CosmoHub has been developed by the Port d'Informaci\'{o} Cient\'{i}fica (PIC), maintained through a collaboration of the Institut de F\'{i}sica d'Altes Energies (IFAE) and the Centro de Investigaciones Energ\'{e}ticas, Medioambientales y Tecnol\'{o}gicas (CIEMAT), and was partially funded by the ``Plan Estatal de Investigaci\'{o}n Cient\'{i}fica y T\'{e}cnica y de Innovaci\'{o}n" program of the Spanish government.

Funding for the SDSS and SDSS-II has been provided by the Alfred P. Sloan Foundation, the Participating Institutions, the National Science Foundation, the U.S. Department of Energy, the National Aeronautics and Space Administration, the Japanese Monbukagakusho, the Max Planck Society, and the Higher Education Funding Council for England. The SDSS Web Site is http://www.sdss.org/. The SDSS is managed by the Astrophysical Research Consortium for the Participating Institutions. The Participating Institutions are the American Museum of Natural History, Astrophysical Institute Potsdam, University of Basel, University of Cambridge, Case Western Reserve University, University of Chicago, Drexel University, Fermilab, the Institute for Advanced Study, the Japan Participation Group, Johns Hopkins University, the Joint Institute for Nuclear Astrophysics, the Kavli Institute for Particle Astrophysics and Cosmology, the Korean Scientist Group, the Chinese Academy of Sciences (LAMOST), Los Alamos National Laboratory, the Max-Planck-Institute for Astronomy (MPIA), the Max-Planck-Institute for Astrophysics (MPA), New Mexico State University, Ohio State University, University of Pittsburgh, University of Portsmouth, Princeton University, the United States Naval Observatory, and the University of Washington.





\bibliographystyle{mnras}
\bibliography{references} 




\appendix
\clearpage

\section{Galaxy clustering statistics and code}
\label{app:code}

We calculate galaxy clustering using the appropriately normalised Landay-Szalay estimator \citep{lsestimator}

\begin{equation}
\xi(r_{\rm p}, \pi) = \frac{DD(r_{\rm p}, \pi) - 2DR(r_{\rm p}, \pi) + RR(r_{\rm p}, \pi)}{RR(r_{\rm p}, \pi)}, 
\end{equation}

DD, DR and RR are normalised Data-Data, Data-Random and Random-Random pair counts. The number of randoms set is always ten times the number of galaxies in a sample, and they are uniformly distributed in the comoving volumes of the samples. $r_{\rm p}$ and $\pi$ are, respectively, the galaxy pair separations transverse and parallel to the line of sight. These separations are defined in terms of the pair of galaxy vectors $\underline{x}_1$ and $\underline{x}_2$

\begin{equation}
\pi = \Bigg| \frac{(\underline{x}_1 - \underline{x}_2).(\underline{x}_1 + \underline{x}_2)}{|\underline{x}_1 + \underline{x}_2|} \Bigg|, 
\end{equation}

\begin{equation}
r_{\rm p} = \sqrt{ (\underline{x}_1 - \underline{x}_2)^2 - \pi^2 }.
\end{equation}

In this analysis we consider only projected galaxy clustering to minimise the impact of the PAUS redshift error. The projected correlation function is given by 

\begin{equation}
\label{eq:wprp}
w_{\rm p}(r_{\rm p}) = 2\int_{0}^{\pi_{\rm max}}\xi(r_{\rm p}, \pi) {\rm d}\pi, 
\end{equation}
where the value of $\pi_{\rm max}$ is a parameter to be set. 

\begin{figure}
 \includegraphics[width=\linewidth, trim={10 10 10 10},clip]{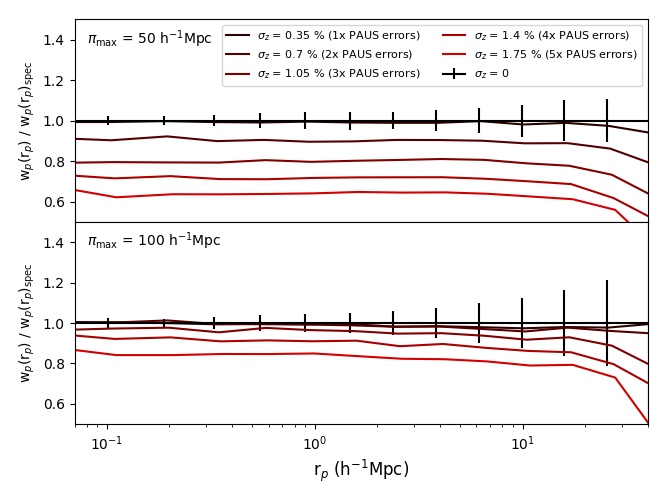}
 \caption{The recovery of the projected galaxy clustering for samples of different Gaussian photometric redshift errors and different values of $\pi_{\rm max}$. Each curve is normalised by the spectroscopic result integrated to the same $\pi_{\rm max}$. The error bars represent the jackknife errors on the spectroscopic result.}
 \label{fig:wprpzerr}
\end{figure}

Fig. \ref{fig:wprpzerr} shows the systematic loss of signal in the projected galaxy clustering for samples with different values of photometric redshift errors relevant to PAUS for two different values of $\pi_{\rm max}$. The sample used was (-$19.5 < \mathrm{M}_{\mathrm{B}}^{\mathrm{h}} < -19.0$) in the redshift range $0.5 < z <  0.63$. The real PAUS data will have a distribution of photometric redshift errors rather than the single Gaussian error assumed here so this plot can inform us on the systematic errors we may introduce for different error distributions. The larger value of $\pi_{\rm max}$ recovers more of the signal but at the cost of increasing the statistical noise. The difference in spectroscopic result between $\pi_{\rm max}$ = 50 and 100h$^{-1}$Mpc is less the 2\%. A value of $\pi_{\rm max}$ of 100 h$^{-1}$Mpc would allow us to use galaxies in the sample with three times the nominal PAUS redshift error and recover the projected clustering within the statistical errors. See \citealt{pablopaper} and \citealt{pablo2014paper} for further discussion on projected correlation recovery in photometric redshift surveys.

All clustering results are calculated using a two point clustering code which is publicly available on github \footnote{ \url{https://github.com/lstothert/two_pcf} }. This is an OpenMP accelerated code which has the ability to calculate monopole and 2D decompositions of the correlation function with flexible linear or logarithmic binning, multiple input/output types and on-the-fly jackknife errors at the expense of very little extra computing time.

The galaxy pairs were binned logarithmically in both $r_{\rm p}$ and $\pi$, which can help reduce the increase in statistical error for large values of $\pi_{max}$.

\section{Clustering samples}
\label{app:samples}

\begin{figure}
  \includegraphics[width=\linewidth]{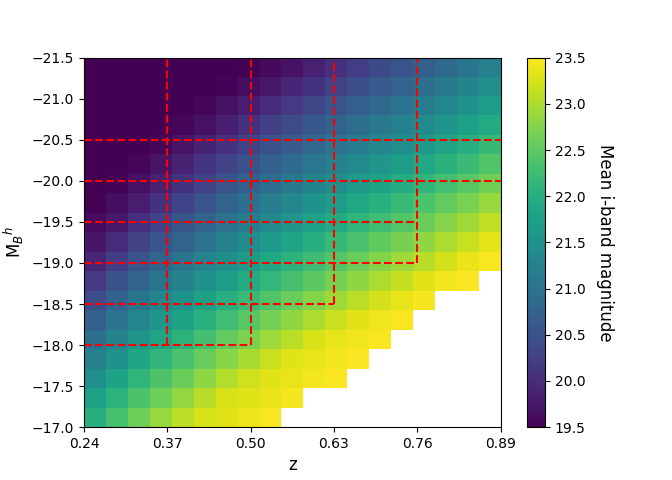}
  \caption{Rest-frame M$_{\rm B}^{\mathrm{h}}$ vs redshift, colour coded by mean $i$-band magnitude for a PAUS mock built to i < 25 without including redshift or photometry errors. The lightcone was built deeper than nominal PAUS depth so as to be certain about the completeness values of the samples. The plot stops at i< 23.5 so the colour gradient through the boxes is more obvious to the reader. The boxes show the sample limits used in the galaxy clustering analysis, chosen to be 99\% complete to $i < 23$ in this lightcone. Note the boxes do not touch the $i = 23$ coloured squares.}
  \label{fig:magz}
\end{figure}

Fig.~\ref{fig:magz} shows the volume limited cuts used to create galaxy clustering samples. The faint limit in M$_{\rm B}^{\mathrm{h}}$ at each redshift was chosen such that the faintest samples were over 99\% complete in a catalogue i < 23 without errors. The scatter in the colour term between the observed i-band and M$_{\rm B}^{\mathrm{h}}$ is responsible for any small amount of incompleteness. The high completeness of the samples can be seen from Fig. \ref{fig:magz} by noting that the bottom right corners of the faintest boxes do not overlap with galaxies with mean i band magnitude of 23. These samples are therefore the samples we would choose if we had perfect photometry, and we then deduce the recoverability of the results when realistic errors are included. The cuts at lower redshift must have a more conservative limit in M$_{\rm B}^{\mathrm{h}}$ than at higher redshift as the scatter between PAUS Blue and the apparent i band magnitude is larger at lower redshift. This is because at the lowest redshift the wavelength difference between the two bands is maximised in the PAUS redshift range so the colour term, and the corresponding colour scatter, is the largest. 

All samples selected along with their completeness and purity once errors are included are listed in Tables~\ref{tab:samplesnocolour} and~\ref{tab:sampleswithcolour}. The definitions of the completeness and purity in those tables can be written as follows. Define N$_{ij}$ as the number of galaxies that lie in sample i in the catalogue without errors and in sample j in the catalogue including errors. Define N$_{i*}$ as the number of galaxies in sample i in the catalogue without errors. Define N$_{*j}$ as the number of galaxies in sample j in the catalogue with errors. The completeness of sample i can now be defined as N$_{ii}$ / N$_{i*}$ and the purity as N$_{ii}$ / N$_{*i}$. Satellite fraction and median halo mass are galaxy weighted quantities. A halo with many satellites may therefore make multiple contributions to the number of satellites and halo masses in a sample. The samples here were split in uniform redshift steps but future work may choose to make the lower redshift bins larger than the higher redshift bins to match the sizes of the volumes probed.

There is high completeness and purity amongst samples split only by redshift and M$_{\rm B}^{\mathrm{h}}$ seen in table \ref{tab:samplesnocolour}, which drops when samples are further split by colour in table \ref{tab:sampleswithcolour}. This shows that the driving source of sample mixing in this work is the colour split. In a fixed luminosity bin the completeness and purity falls with redshift as the photometry errors are larger for apparently fainter samples. This also holds once galaxies are split by colour.

The number density of the brightest galaxies increases with increasing redshift as the star formation rate of the universe, and therefore the amplitude of the M$_{\rm B}^{\mathrm{h}}$ luminosity function, increases with redshift. These trends are also seen in fainter samples but aren't as clear once errors are included. Brighter galaxies live in larger halos and this trend is particularly strong for red galaxies. These red galaxies also on average live in significantly larger halos than their blue counterparts with the same luminosity and redshift. At fixed colour and luminosity the median halo mass increases with decreasing redshift as the dark matter growth rate is large on small non-linear scales over this redshift range. 

\begin{table*}
  \begin{center}
    
 \begin{tabular}{ccccccccc}
 \hline 
 z-min & z-max  & M$_{\rm B}^{\mathrm{h}}$ bright & M$_{\rm B}^{\mathrm{h}}$ faint & Comp (\%) & Purity (\%) & $\overline{n}$ & Sat frac & Median M$_{\rm halo}$\\ 
  Volume & $(10^{6} h^{-3} {\rm Mpc^{-3}} )$ &  &  &  &  & $(10^{-3} h^{-3} {\rm Mpc^{-3}} )$ &  & $(10^{11}  h^{-1} {\rm M_{\odot}})$\\ 
   
 \hline
   0.24 & 0.37 & -18.5 & -18.0 & 89.6 & 88.6 & 7.51 & 0.273 & 2.43 \\
4.626 &  & -19.0 & -18.5 & 92.4 & 91.9 & 6.22 & 0.259 & 3.58 \\
 &  & -19.5 & -19.0 & 94.6 & 93.5 & 4.89 & 0.221 & 4.58 \\
 &  & -20.0 & -19.5 & 95.4 & 94.7 & 3.41 & 0.153 & 5.48 \\
 &  & -20.5 & -20.0 & 96.2 & 95.4 & 1.8 & 0.1 & 6.22 \\
 &  & None & -20.5 & 97.3 & 96.2 & 1.02 & 0.073 & 8.66 \\
\hline
0.37 & 0.5 & -18.5 & -18.0 & 81.9 & 81.7 & 8.14 & 0.291 & 2.34 \\
8.262 &  & -19.0 & -18.5 & 87.5 & 86.9 & 6.73 & 0.275 & 3.53 \\
 &  & -19.5 & -19.0 & 90.8 & 90.8 & 5.44 & 0.242 & 4.6 \\
 &  & -20.0 & -19.5 & 93.2 & 92.9 & 3.88 & 0.179 & 5.36 \\
 &  & -20.5 & -20.0 & 94.1 & 94.2 & 2.1 & 0.113 & 6.05 \\
 &  & None & -20.5 & 96.1 & 96.1 & 1.22 & 0.079 & 8.79 \\
\hline
0.5 & 0.63 & -19.0 & -18.5 & 81.2 & 82.0 & 6.22 & 0.273 & 3.31 \\
12.22 &  & -19.5 & -19.0 & 87.0 & 86.2 & 5.31 & 0.234 & 4.28 \\
 &  & -20.0 & -19.5 & 90.2 & 89.9 & 3.91 & 0.174 & 5.03 \\
 &  & -20.5 & -20.0 & 92.2 & 91.7 & 2.15 & 0.117 & 5.78 \\
 &  & None & -20.5 & 95.2 & 95.0 & 1.25 & 0.076 & 8.26 \\
\hline
0.63 & 0.76 & -19.5 & -19.0 & 82.2 & 83.7 & 4.96 & 0.232 & 4.09 \\
16.177 &  & -20.0 & -19.5 & 86.9 & 86.7 & 3.91 & 0.177 & 4.85 \\
 &  & -20.5 & -20.0 & 89.9 & 89.6 & 2.23 & 0.118 & 5.5 \\
 &  & None & -20.5 & 94.1 & 93.9 & 1.3 & 0.08 & 8.08 \\
\hline
0.76 & 0.89 & -20.5 & -20.0 & 87.9 & 87.4 & 2.62 & 0.129 & 5.42 \\
19.922 &  & None & -20.5 & 93.1 & 92.9 & 1.53 & 0.083 & 7.99 \\

\end{tabular}

 \caption{Table of galaxy clustering samples used in this analysis. Completeness, purity and satellite fraction are defined in the text. $\overline{n}$ is the number density of the sample.}
 \label{tab:samplesnocolour}

\end{center}
\end{table*}

\begin{table*}
  \begin{center}
    
 \begin{tabular}{cccccccccc}
 \hline 
 
 z-min & z-max  & Colour & M$_{\rm B}^{\mathrm{h}}$ bright & M$_{\rm B}^{\mathrm{h}}$ faint & Comp (\%) & Purity (\%) & $\overline{n}$ & Sat frac & Median M$_{\rm halo}$\\ 
   Volume & $(10^{6} h^{-3} {\rm Mpc^{-3}} )$ &  & &   &  &  & $(10^{-3} h^{-3} {\rm Mpc^{-3}} )$ &  & $(10^{11}  h^{-1} {\rm M_{\odot}})$\\ 

\hline
0.24 & 0.37 & red & -18.5 & -18.0 & 73.9 & 63.9 & 3.43 & 0.481 & 13.3 \\
4.626 &  &  & -19.0 & -18.5 & 85.8 & 80.2 & 3.11 & 0.47 & 19.4 \\
 &  &  & -19.5 & -19.0 & 92.2 & 88.2 & 2.52 & 0.401 & 20.4 \\
 &  &  & -20.0 & -19.5 & 93.5 & 91.6 & 1.68 & 0.286 & 20.9 \\
 &  &  & -20.5 & -20.0 & 94.9 & 92.8 & 0.788 & 0.193 & 23.2 \\
 &  &  & None & -20.5 & 96.6 & 94.6 & 0.454 & 0.13 & 64.7 \\
 &  & blue & -18.5 & -18.0 & 71.7 & 78.3 & 4.08 & 0.097 & 1.91 \\
 &  &  & -19.0 & -18.5 & 81.8 & 86.2 & 3.11 & 0.048 & 2.36 \\
 &  &  & -19.5 & -19.0 & 88.9 & 91.0 & 2.37 & 0.029 & 2.89 \\
 &  &  & -20.0 & -19.5 & 92.2 & 92.6 & 1.73 & 0.023 & 3.63 \\
 &  &  & -20.5 & -20.0 & 94.0 & 94.3 & 1.01 & 0.026 & 4.69 \\
 &  &  & None & -20.5 & 95.9 & 95.5 & 0.571 & 0.028 & 6.37 \\
\hline
0.37 & 0.5 & red & -18.5 & -18.0 & 56.3 & 46.3 & 3.94 & 0.412 & 6.61 \\
8.262 &  &  & -19.0 & -18.5 & 69.5 & 63.6 & 3.4 & 0.426 & 12.2 \\
 &  &  & -19.5 & -19.0 & 81.1 & 76.9 & 2.88 & 0.401 & 16.7 \\
 &  &  & -20.0 & -19.5 & 88.2 & 84.0 & 2 & 0.31 & 17.9 \\
 &  &  & -20.5 & -20.0 & 91.2 & 88.5 & 0.964 & 0.212 & 22.2 \\
 &  &  & None & -20.5 & 94.1 & 93.9 & 0.571 & 0.136 & 50.7 \\
 &  & blue & -18.5 & -18.0 & 55.2 & 64.2 & 4.2 & 0.178 & 1.95 \\
 &  &  & -19.0 & -18.5 & 66.3 & 71.1 & 3.33 & 0.12 & 2.4 \\
 &  &  & -19.5 & -19.0 & 76.3 & 81.0 & 2.55 & 0.062 & 2.88 \\
 &  &  & -20.0 & -19.5 & 84.5 & 88.3 & 1.88 & 0.04 & 3.51 \\
 &  &  & -20.5 & -20.0 & 89.3 & 91.7 & 1.14 & 0.029 & 4.41 \\
 &  &  & None & -20.5 & 94.2 & 94.4 & 0.651 & 0.028 & 6.02 \\
\hline
0.5 & 0.63 & red & -19.0 & -18.5 & 64.2 & 61.0 & 2.94 & 0.448 & 12.7 \\
12.22 &  &  & -19.5 & -19.0 & 74.9 & 73.5 & 2.58 & 0.4 & 14.9 \\
 &  &  & -20.0 & -19.5 & 83.9 & 82.8 & 1.86 & 0.316 & 16.6 \\
 &  &  & -20.5 & -20.0 & 88.8 & 87.7 & 0.94 & 0.219 & 18.1 \\
 &  &  & None & -20.5 & 93.4 & 93.5 & 0.579 & 0.129 & 36.4 \\
 &  & blue & -19.0 & -18.5 & 65.2 & 69.5 & 3.27 & 0.114 & 2.29 \\
 &  &  & -19.5 & -19.0 & 75.6 & 75.7 & 2.73 & 0.076 & 2.8 \\
 &  &  & -20.0 & -19.5 & 84.1 & 84.8 & 2.05 & 0.044 & 3.35 \\
 &  &  & -20.5 & -20.0 & 89.5 & 89.5 & 1.21 & 0.038 & 4.17 \\
 &  &  & None & -20.5 & 94.1 & 93.7 & 0.67 & 0.03 & 5.78 \\
\hline
0.63 & 0.76 & red & -19.5 & -19.0 & 66.5 & 65.2 & 2.42 & 0.378 & 11.6 \\
16.177 &  &  & -20.0 & -19.5 & 76.1 & 73.6 & 1.88 & 0.302 & 12.9 \\
 &  &  & -20.5 & -20.0 & 83.4 & 81.4 & 0.989 & 0.214 & 14.8 \\
 &  &  & None & -20.5 & 91.3 & 91.3 & 0.61 & 0.132 & 27.7 \\
 &  & blue & -19.5 & -19.0 & 67.0 & 70.6 & 2.54 & 0.093 & 2.77 \\
 &  &  & -20.0 & -19.5 & 76.2 & 78.3 & 2.02 & 0.06 & 3.32 \\
 &  &  & -20.5 & -20.0 & 84.4 & 85.5 & 1.25 & 0.041 & 4.01 \\
 &  &  & None & -20.5 & 91.7 & 91.5 & 0.686 & 0.034 & 5.56 \\
\hline
0.76 & 0.89 & red & -20.5 & -20.0 & 77.8 & 74.7 & 1.22 & 0.214 & 12.1 \\
19.922 &  &  & None & -20.5 & 88.7 & 87.9 & 0.735 & 0.129 & 22.6 \\
 &  & blue & -20.5 & -20.0 & 78.5 & 80.4 & 1.41 & 0.054 & 3.88 \\
 &  &  & None & -20.5 & 88.5 & 88.9 & 0.795 & 0.041 & 5.39 \\

\end{tabular}

\caption{Table of galaxy clustering samples used in this analysis including colour splits. Completeness, purity and satellite fraction are defined in the text. $\overline{n}$ is the number density of the sample.}
 \label{tab:sampleswithcolour}

\end{center}
\end{table*}


\bsp	

\label{lastpage}
\end{document}